\documentclass[twocolumn]{aastex62}

\usepackage{amsmath}
\usepackage{natbib}
\usepackage{multirow}
\usepackage{hyperref}
\usepackage{bm}

\begin{document}

\title{Analysis of the duration--hardness ratio plane of gamma-ray bursts with skewed distributions}

\author{Mariusz Tarnopolski}
\email{mariusz.tarnopolski@uj.edu.pl}

\affiliation{Astronomical Observatory, Jagiellonian University, Orla 171, 30--244, Krak\'ow, Poland}

\begin{abstract}
It was recently shown that the $T_{90}-H_{32}$ distributions of gamma-ray bursts from {\it CGRO}/BATSE and {\it Fermi}/GBM are well described by a mixture of only two skewed components, making the presumed third, intermediate class unnecesary. The {\it Swift}/BAT, {\it Konus}-Wind, {\it RHESSI} and {\it Suzaku}/WAM data sets are found to be consistent with a two-class description as well.
\end{abstract}


\section{Introduction}

The two widely accepted classes of gamma-ray bursts (GRBs), short and long, are with confidence ascribed to mergers of compact objects and collapse of massive stars, respectively. A third, intermediate class \citep{horvath98}, remains putative. Its existence was claimed based on univariate and bivariate analyses of GRB observables modeled with Gaussian distributions \citep{mukh,horvath02,horvath08,zhang08,huja,ripa09,horvath10,veres,zitouni,zhang16,horvath18}, but also has been put into doubt several times \citep{bystricky,ripa12,tarnopolski15b,zitouni,bhat,tarnopolski16c,tarnopolski16b,ohmori,yang,kulkarni,zitouni18}. Gaussian models, however, may not be the appropriate approach\footnote{\citet{mukh} noted that ''the distributions often seem bimodal with asymmetrical non-Gaussian shapes'', but failed to employ skewed distributions in modeling and proceeded considering multinormal distributions.} \citep{koen,tarnopolski15b,koen17}, as it has been already shown that the univariate distributions of $T_{90}$ \citep{tarnopolski16a,tarnopolski16c,kwong} and bivariate $T_{90}-H_{32}$ ones \citep{tarnopolski19} are better described by mixtures of two skewed components rather than three Gaussian ones. In this work the $T_{90}-H_{32}$ plane is examined in case of data sets from four other satellites: {\it Swift}/BAT, {\it Konus}-Wind, {\it RHESSI}, and {\it Suzaku}/WAM.

\vspace{1cm}
\section{Data}

The following data sets are investigated: 1033 GRBs from the {\it Swift}/BAT catalogue \citep{lien16}, 1143 GRBs observed by {\it Konus}-Wind \citep{svinkin16}, 427 GRBs detected by {\it RHESSI} \citep{ripa09}, and 259 GRBs from {\it Suzaku}/WAM \citep{ohmori}. The bivariate distributions of duration $T_{90}$ and hardness ratio $H_{32}$ in the log-log plane are examined. For each instrument, fluences $F$ in different energy bands are available, hence the definitions of $H_{32}$ are: $H_{32}=\frac{F_{50-100\,{\rm keV}}}{F_{25-50\,{\rm keV}}}$ for {\it Swift}; $H_{32}=\frac{F_{200-750\,{\rm keV}}}{F_{50-200\,{\rm keV}}}$ for {\it Konus}; $H_{32}=\frac{F_{120-1500\,{\rm keV}}}{F_{25-120\,{\rm keV}}}$ for {\it RHESSI}; and $H_{32}=\frac{F_{240-520\,{\rm keV}}}{F_{110-240\,{\rm keV}}}$ for {\it Suzaku}. 

\section{Methodology}

The methodology is the same as in \citep{tarnopolski19}. Two- and three-component mixtures of the following bivariate distributions are fitted: regular Gaussian (2G and 3G), skew-normal (2SN and 3SN), Student $t$ (2T and 3T), and skew-Student (2ST and 3ST).  The fits are compared using the small sample Akaike \citep{hurvich89} and Bayesian Information Criteria ($AIC_c$ and $BIC$). $AIC_c$ is liberal, and has a tendency to overfit. $BIC$ is much more stringent, and tends to underfit. Therefore, when the two $IC$ point at different models, the truth lies somewhere in between. (See \citealt{tarnopolski19} for details.) The fitting is performed using the {\sc R} package \texttt{mixsmsn}\footnote{\url{https://cran.r-project.org/web/packages/mixsmsn/index.html}} \citep{prates}.

\vspace{1cm}
\section{Results}

\begin{figure}[t!]
\resizebox{\hsize}{!}{\includegraphics[clip=true]{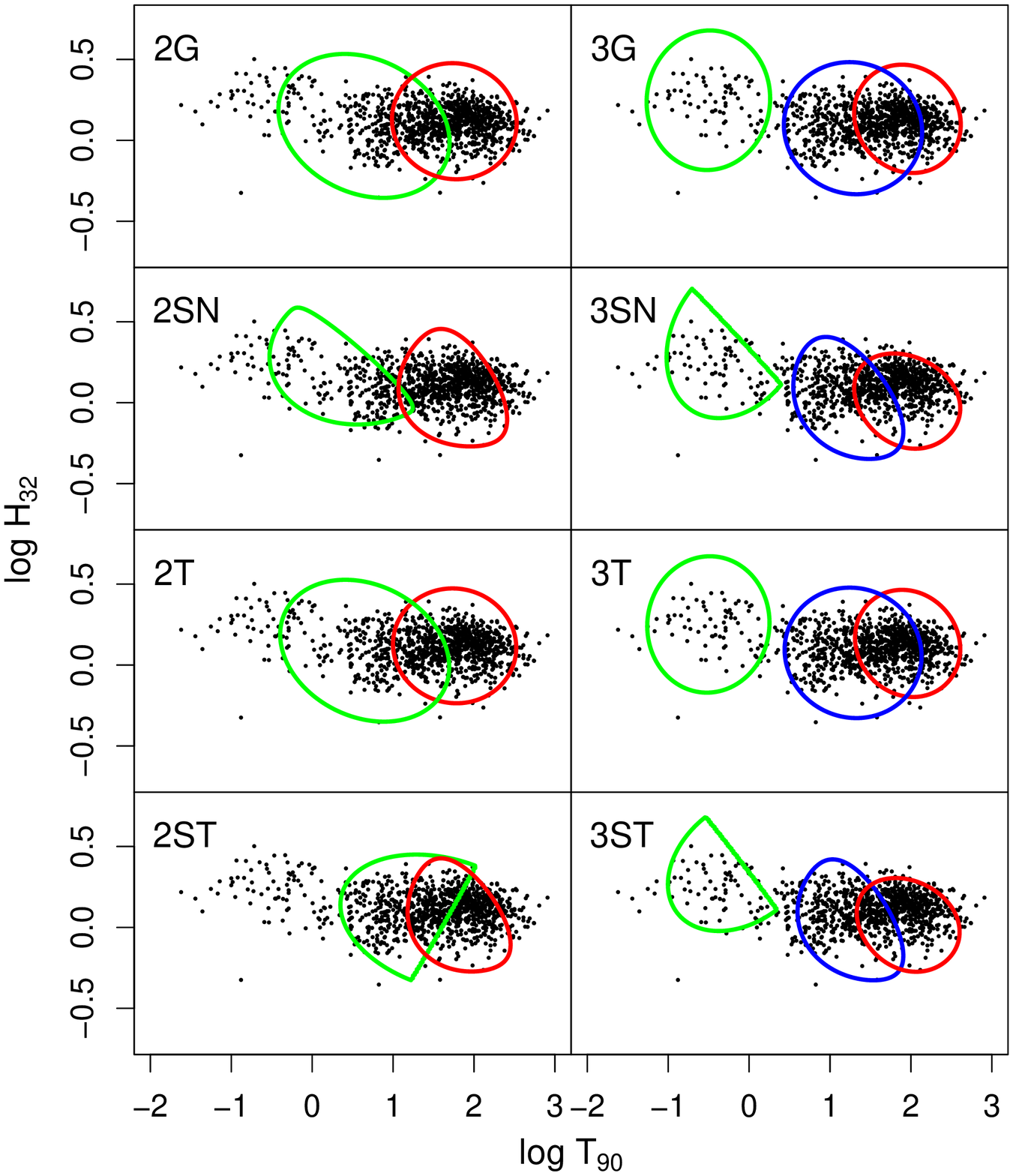}}
\resizebox{\hsize}{!}{\includegraphics[clip=true]{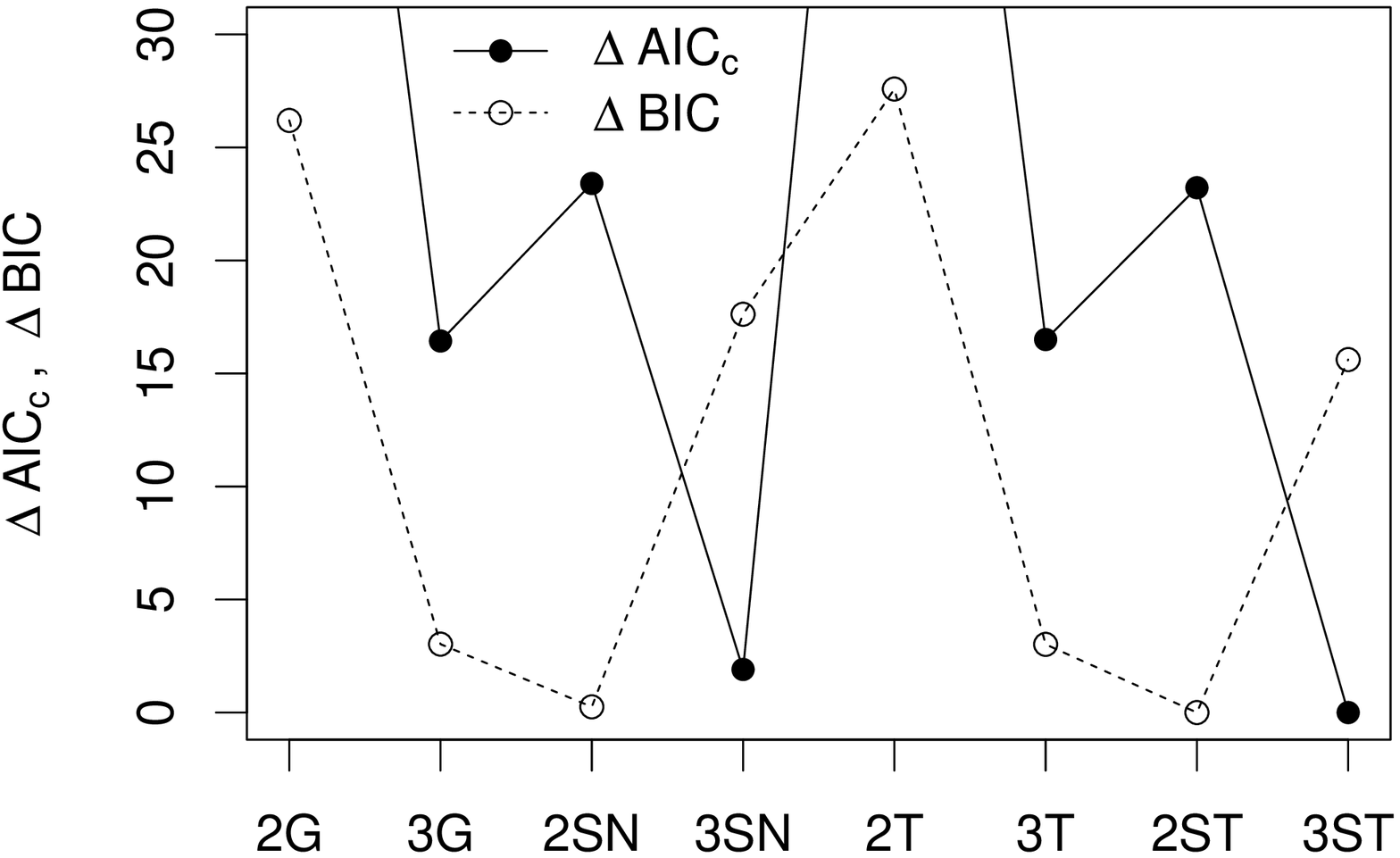}}
\caption{
\footnotesize
Fittings and $\Delta IC$ scores for {\it Swift} GRBs.
}
\label{fig1}
\end{figure}

\begin{figure}[t!]
\resizebox{\hsize}{!}{\includegraphics[clip=true]{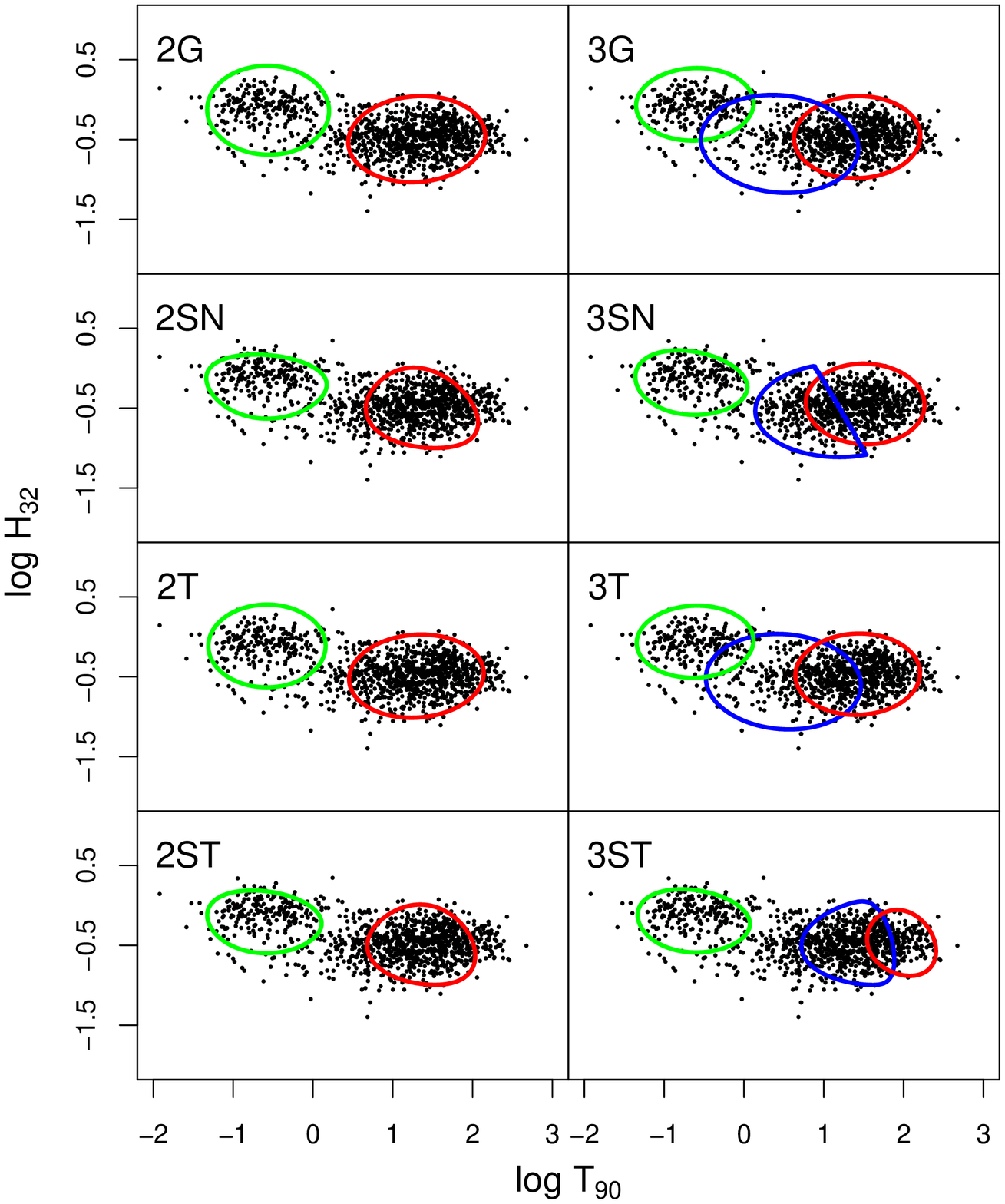}}
\resizebox{\hsize}{!}{\includegraphics[clip=true]{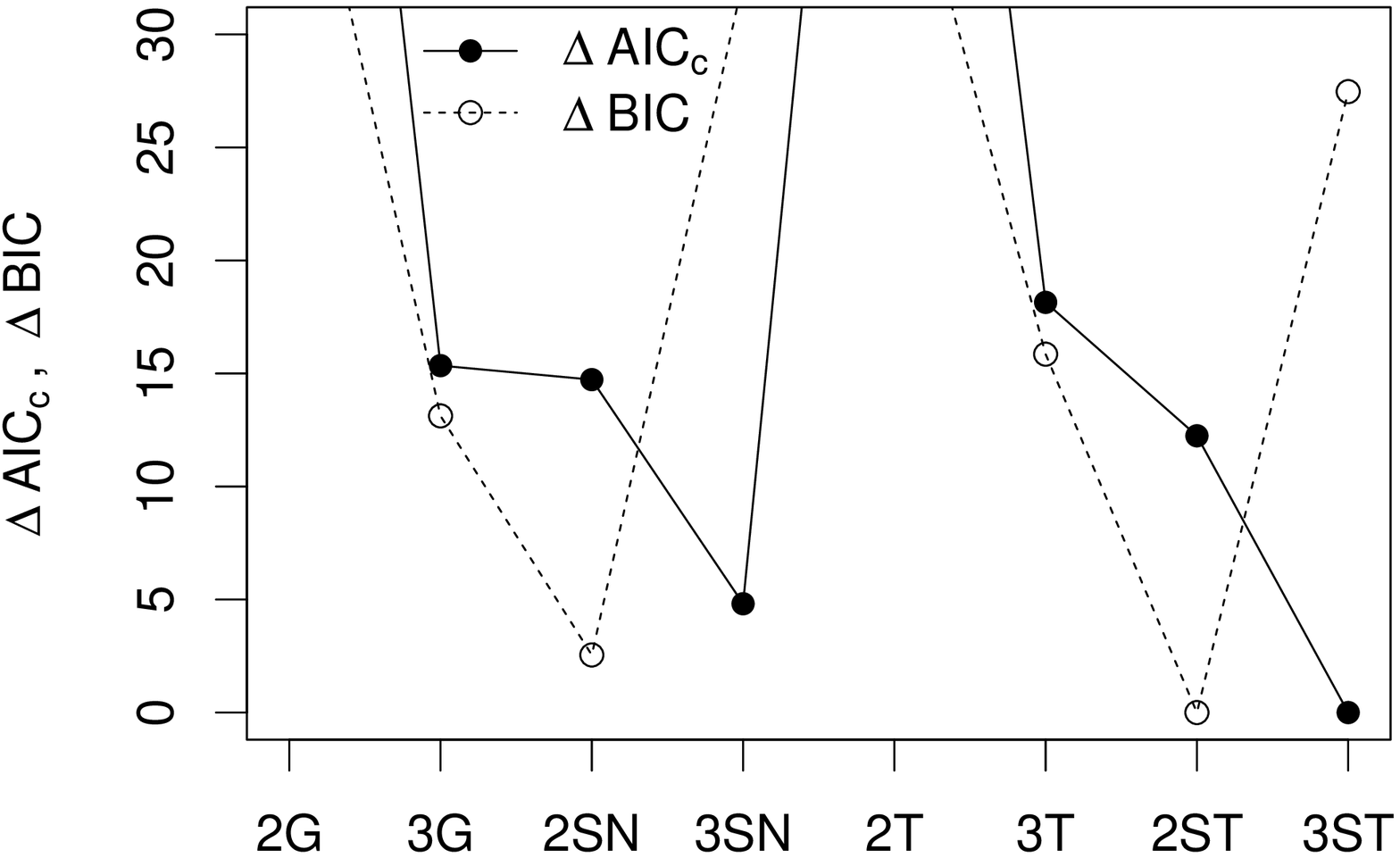}}
\caption{
\footnotesize
Fittings and $\Delta IC$ scores for {\it Konus} GRBs.
}
\label{fig2}
\end{figure}

\begin{figure}[t!]
\resizebox{\hsize}{!}{\includegraphics[clip=true]{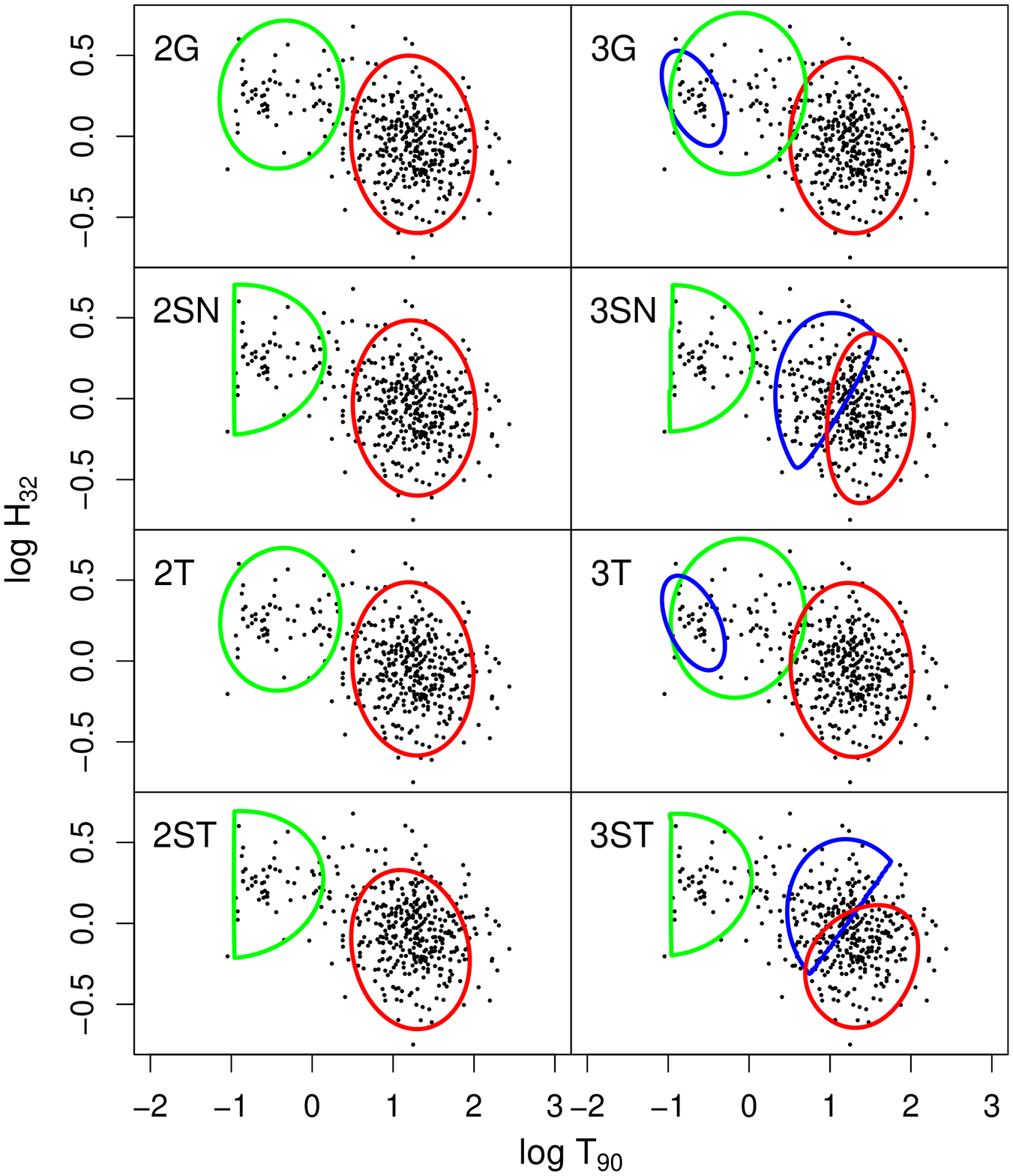}}
\resizebox{\hsize}{!}{\includegraphics[clip=true]{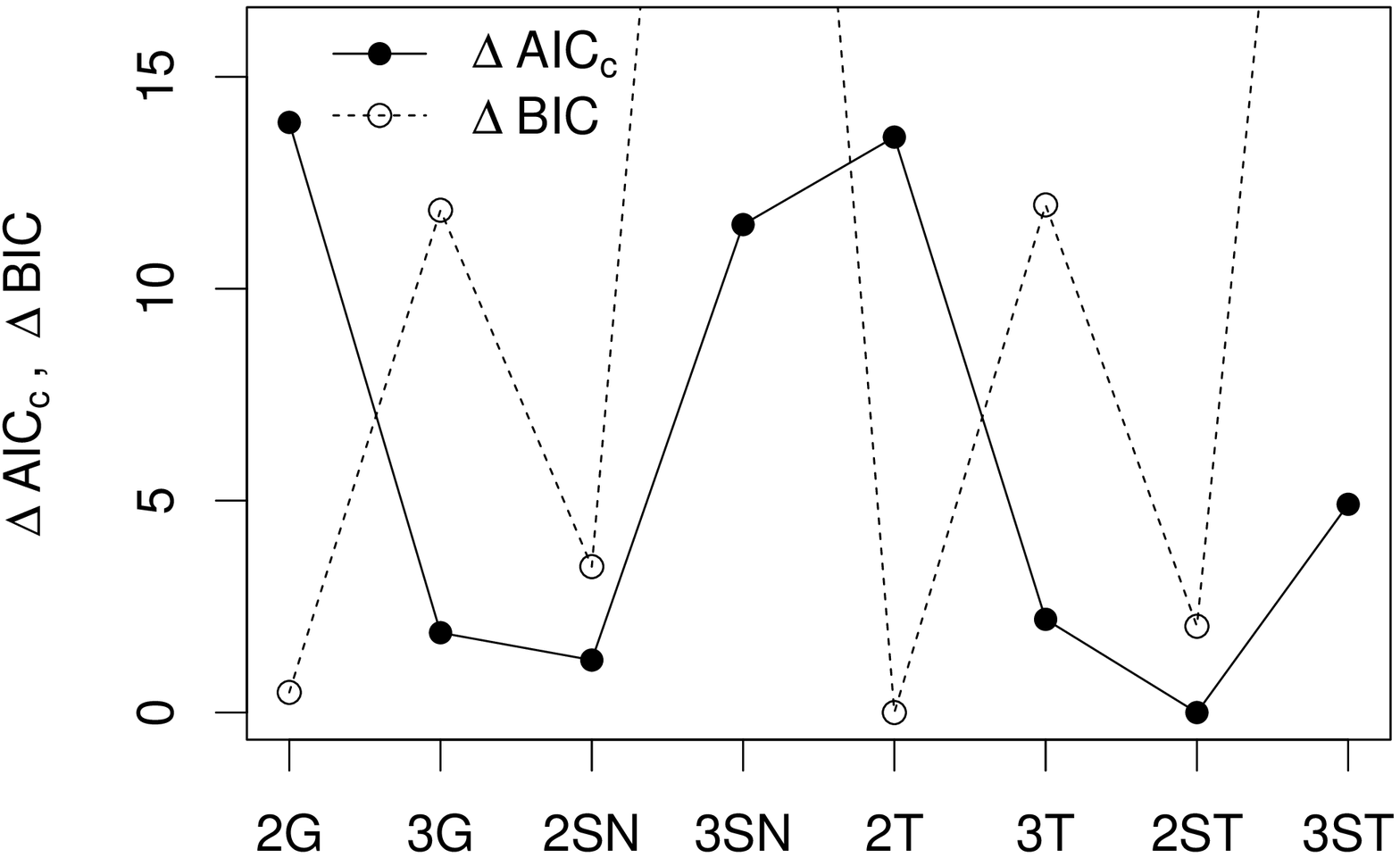}}
\caption{
\footnotesize
Fittings and $\Delta IC$ scores for {\it RHESSI} GRBs.
}
\label{fig3}
\end{figure}

\begin{figure}[t!]
\resizebox{\hsize}{!}{\includegraphics[clip=true]{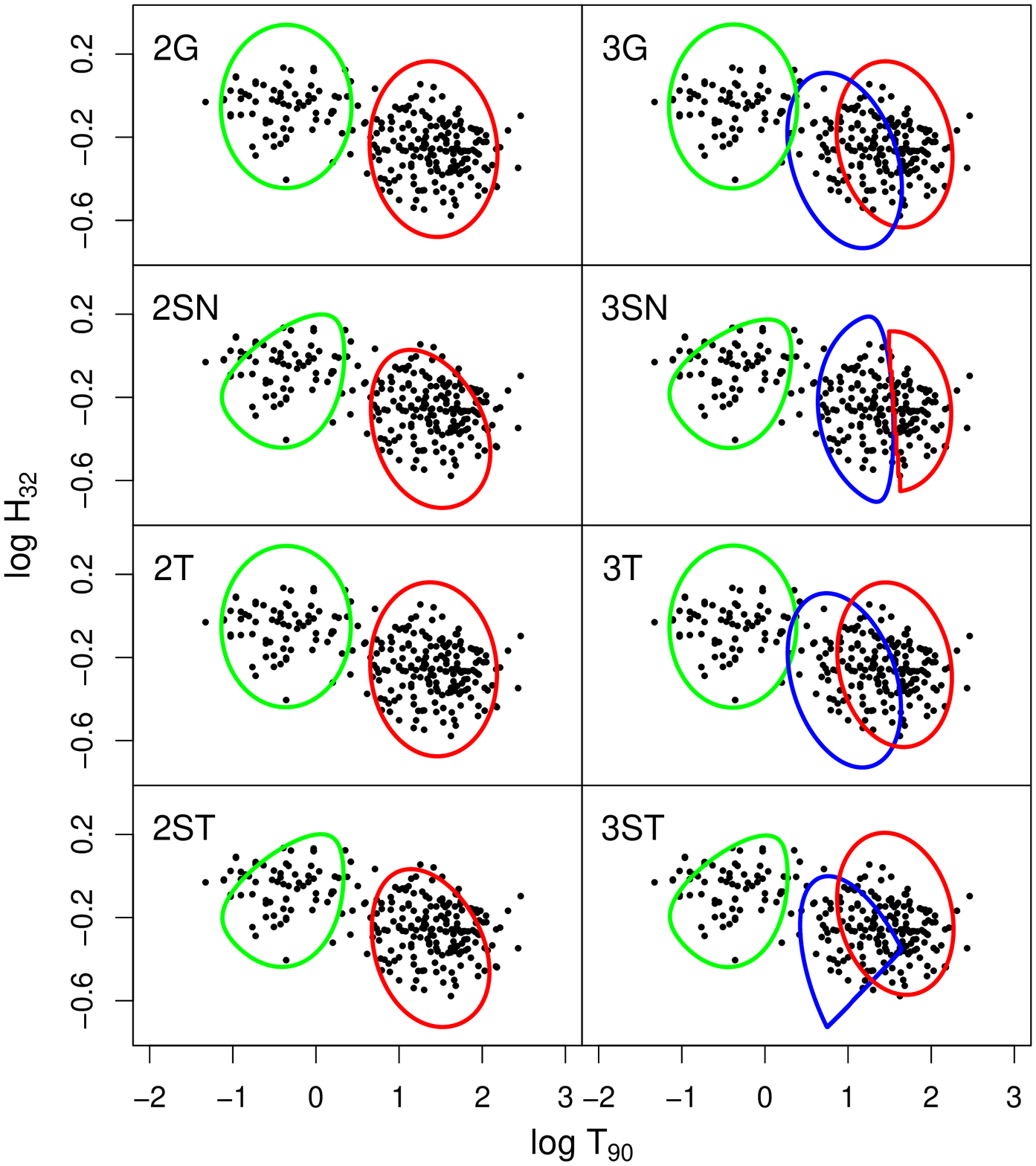}}
\resizebox{\hsize}{!}{\includegraphics[clip=true]{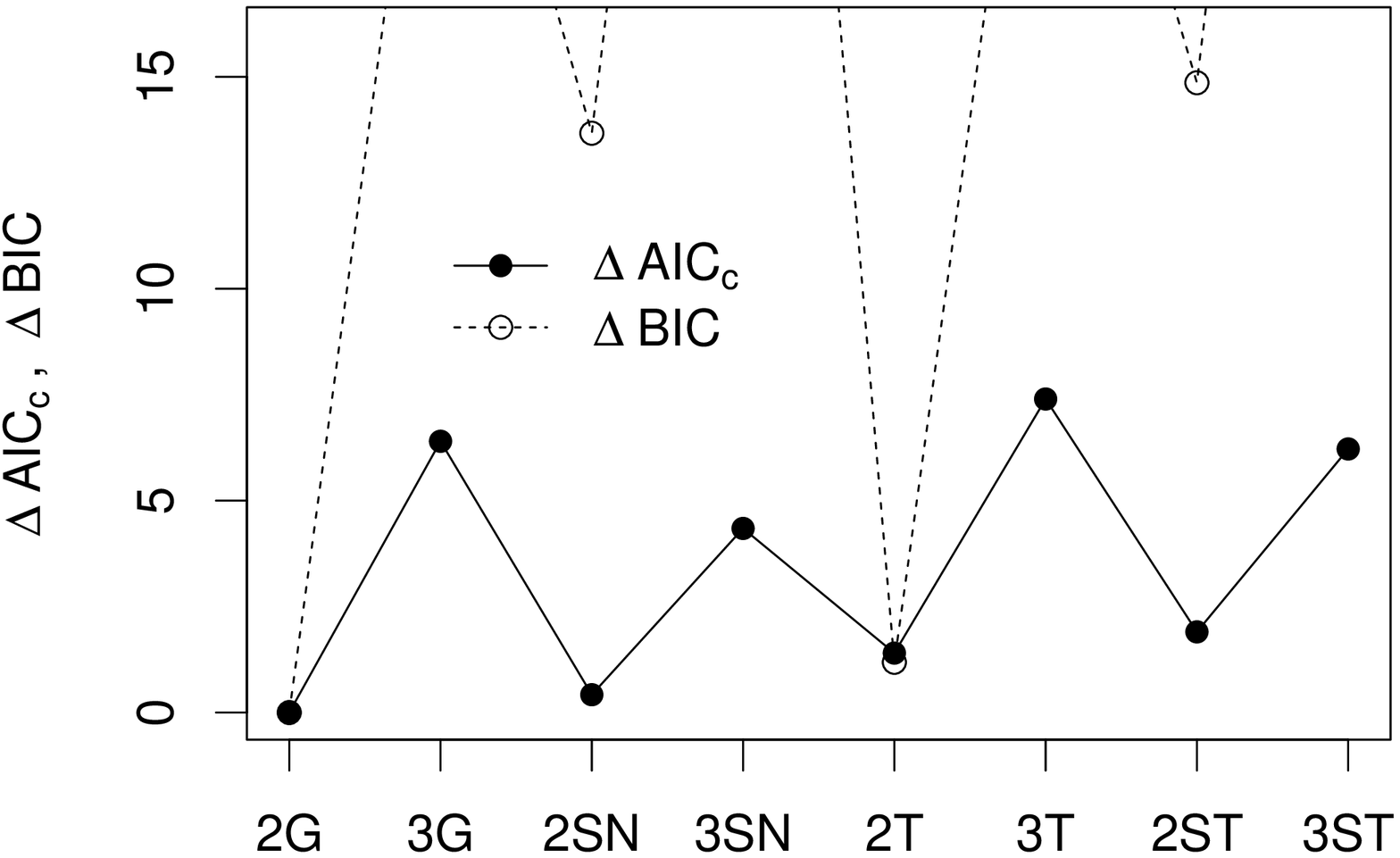}}
\caption{
\footnotesize
Fittings and $\Delta IC$ scores for {\it Suzaku} GRBs.
}
\label{fig4}
\end{figure}

The results are displayed in Figs.~\ref{fig1}--\ref{fig4}. For {\it Swift} and {\it Konus} no clear answer is obtained, however both $IC$ point at skewed distributions (see bottom panels of Figs.~\ref{fig1} and \ref{fig2}). For {\it Swift}, the $BIC$ yields 2ST and 2SN, while $AIC_c$ gives 3ST and 3SN. Henceforth, the lack of a third component in the data cannot be confidently ruled out; on the other hand, its presence is also not unambiguously supported. {\it Konus} gives remarkably similar results.

In case of {\it RHESSI} (see Fig.~\ref{fig3}), both $IC$ point unequivocally at 2-component mixtures, however $BIC$ prefers symmetric distributions (2G and 2T), while $AIC_c$ hints at skewed ones (2ST and 2SN). Suzaku, the smallest data set examined, can be with no doubt well modeled with only 2 components, with 2G being the simplest model (see Fig.~\ref{fig4}).

\section{Discussion}

GRBs from BATSE and {\it Fermi} can be confidently divided into only two classical groups, short and long; the elusive soft-intermediate class is not necessary to satisfactorily describe the data \citep{tarnopolski15b,tarnopolski16a,tarnopolski16c,tarnopolski19}. In case of {\it Swift} and {\it Konus}, however, no firm conclusion can be formulated---the $IC$ point at either two or three classes. The smallest data sets---{\it RHESSI} and {\it Suzaku}---can be adequately construed as consisting of two groups, although due to the smallness of these samples, the more subtle structure in the $T_{90}-H_{32}$ plane can simply be not traced prominently enough.

The asymmetry of the data, manifested via skewed distributions, might come from a non-symmetric distribution of the envelope masses of the progenitors of the long GRBs or other inherently asymmetrical distributions of physical parameters governing the progenitors or GRBs themselves; from the impact of the redshift distribution on the observables; or a combination of the listed possibilities \citep{tarnopolski15b,zitouni,tarnopolski16a,tarnopolski16c,tarnopolski16b,tarnopolski19}.

\section{Conclusions}

No definite signs of the putative third GRB class are visible in the examined data. On the other hand, the {\it Swift} and {\it Konus} data yield inconclusive. It is desirable to have the exact shape of the observed distributions derived from a physical theory, or inferred on the grounds of statistics, which has not been convincingly realized thus far.
 
\begin{acknowledgements}
The author is grateful to P{\'e}ter Veres for the {\it Fermi} data, Dmitry Svinkin for the {\it Konus} data, Norisuke Ohmori for the {\it Suzaku} data, Amy Lien for help with the {\it Swift} data, and Jakub {\v R}{\'{\i}}pa and Natalia \.Zywucka for discussions. Support by the Polish National Science Center through an OPUS grant No. 2017/25/B/ST9/01208 is acknowledged.
\end{acknowledgements}

\newpage

\bibliography{bibliography}

\end{document}